\documentclass[aps,prl,gif,twocolumn]{revtex4}
\usepackage{graphicx,epsf}
\usepackage{dcolumn}
\usepackage{bm}

\begin{document}


\def\appConv{Appendix~A}
\def\appBeer{Appendix~B}
\def\appHValid{Appendix~C}

\def\HfO2{{HfO$_2$}}
\def\SiO2{{SiO$_2$}}
\def\Al2O3{{Al$_2$O$_3$}}
\def\k{{$\kappa$~}}
\def\ks{{$\kappa_s$~}}
\def\ksten{{$\tensor{\kappa}_s$~}}
\def\kinf{{$\kappa_\infty$~}}
\def\kion{{$\kappa_{ion}$~}}
\def\kionten{{$\tensor{\kappa}_{ion}$~}}
\def\kinften{{$\tensor{\kappa}_\infty$~}}
\def\cm1{{cm$^{-1}$}}
\def\zstar{{$Z^{\star}$~}}

\def\comment#1{{\large\textsl{#1}}}
\def\degree {{$^\circ$}}
\def\degrees{{$^\circ$}}
\def\eq#1{{Eq.~(\ref{eq:#1})}}
\def\fig#1{{Fig.~\ref{fig:#1}}}
\def\sec#1{{Sec.~\ref{sec:#1}}}
\def\inv{^{-1}}
\def\micron {\hbox{$\mu$m}}
\def\microns{\micron}
\def\Ref#1{{Ref.~\onlinecite{#1}}}  
\def\tab#1{{Table~\ref{tab:#1}}}
\def\tauOne{\tau^{(1)}}
\def\tVec{\hbox{\bf t}}
\def\thetaDet{\theta_{DET}}

\def\qvec{{\vec q}}
\def\pvec{{\vec p}}
\def\Avec{{\vec A}}
\def\qhat{{\hat q}}
\def\qperphat{{\hat q_\perp}}
\def\ekpq{{E_{\kvec+\qvec}}}
\def\ek{{E_{\kvec}}}
\def\Omegabar{{\bar\Omega}}
\def\omegabar{{\bar\omega}}
\def\omegap{{\omega_p}}
\def\kf{{k_F}}
\def\kappaf{{\kappa_F}}
\def\mone{{-1}}
\def\re{{\rm{Re\,}}}
\def\im{{\rm{Im\,}}}
\def\twopi{{2 \pi}}
\def\wpm{w_\pm}
\def\FWlindhard{Appendix~A}
\def\lindhardTrans{Appendix~B}
\def\ftr{{f^{tr}}}
\def\PN{Pines and Nozi{\`e}res}
\def\Bohm{B{\"o}hm}
\def\Nifosi{Nifos{\'\i}}
\def\prin{{\cal P}}

\def\imagOmegaSq{the Appendix}
\def\epsTensor{{\buildrel \leftrightarrow \over \epsilon}}
\def\chiTensor{{\buildrel \leftrightarrow \over \chi}}
\def\idenTensor{{\buildrel \leftrightarrow \over I}}
\def\epsTrans{{\epsilon^{(t)}}}
\def\epsLong{{\epsilon^{(\ell)}}}
\def\epsTransInv{{\epsilon^{(t)-1}}}
\def\epsLongInv{{\epsilon^{(\ell)-1}}}

\def\MvecA{{M^{(\vec A)}}}
\def\MdivA{{M^{(\nabla \cdot \vec A)}}}
\def\Mphi{{M^{(\phi)}}}
\def\backGrad{{\buildrel \leftarrow \over \nabla}}
\def\backMom{i \hbar \backGrad}

\def\half{{1/2}}
\def\minusHalf{{-1/2}}
\def\threeHalves{{3/2}}
\def\minusThreeHalves{{-3/2}}

\newenvironment{bulletList}{\begin{list}{$\bullet$}{}}{\end{list}}

\title{Graphing and Grafting Graphene: Classifying Finite Topological Defects}

\author{Eric Cockayne}

\affiliation{Ceramics Division, Material Measurement
Laboratory, National Institute of Standards and
Technology, Gaithersburg, Maryland 20899 USA; Electronic
address: eric.cockayne@nist.gov}
 

\begin{abstract}
\centering
 The structure of finite-area topological defects in graphene is described
in terms of both the direct honeycomb lattice and its dual
triangular lattice.  Such defects are equivalent to cutting out a 
patch of graphene and replacing it with a different patch 
with the same number of dangling bonds.  
An important subset of these defects, bound by a closed loop of 
alternating 5- and 7-membered carbon rings, explains most finite-area
topological defects that have been experimentally observed.  
Previously unidentified defects seen in scanning tunneling microscope (STM) 
images of graphene grown on SiC are identified as isolated divacancies or 
divacancy clusters.


\end{abstract}


\maketitle
\thispagestyle{empty}


Producing commercial graphene-based devices will require the ability
to grow large sheets of high-quality graphene.  Several techniques
for producing graphene exist, including mechanical exfoliation from
graphite\cite{Novoselov05}, chemical exfoliation from
graphite\cite{Hernandez08}, chemical reduction of graphene
oxide\cite{Stankovich07}, segregation of carbon
from metal crystals\cite{Yu08}, chemical vapor deposition of C onto
metal surfaces\cite{Kim09,Li09}, and thermal desorption of Si from 
SiC\cite{Rutter07,Kedzierski08}.
Graphene produced by the above methods is often found to contain defects,
such as vacancies\cite{Ugeda10} or grain 
boundaries\cite{Coraux08,Wofford10,Huang11}.
Defects decrease the high mobility of graphene\cite{Chen09}; therefore,
it is desirable to reduce or eliminate the number of defects.
Conversely, one may desire to tune the properties of graphitic
materials, {\it e.g.} the bandgap, by intentionally creating
and manipulating defect 
structures\cite{Lusk08,Lahiri10,Lusk10,Terrones10}.
Toward either end, it is necessary to classify the kinds of defects that form
in graphene and to correlate them with the growth conditions.

One class of defects, frequently observed in scanning tunneling microscope (STM) images of 
ultrahigh-vacuum graphene growth via Si desorption from SiC\cite{Rutter07,Guisinger08}, 
exhibits regions, roughly several nm in size, of strongly 
perturbed electronic structure completely surrounded by ordered graphene.  
The finite range of the electronic structure perturbation suggests that 
these defects could be created
or healed by the motion of a relatively small numbers of atoms, and may
therefore be among the most important defects in graphene.
We recently identified\cite{Cockayne11} the sixfold symmetric defect seen
in Ref.~\onlinecite{Rutter07} as the ``flower" 
defect\cite{Park10,Meyer11} (shown below),
a topological defect that can be described as the rotation of 24 central atoms
in ideal graphene by 30$^{\circ}$. 
Other finite-area defects, with twofold, and threefold symmetry,
are seen in Ref.~\onlinecite{Rutter07}, but their structures have not been previously 
identified\cite{Guisinger08}.  
In this paper, we describe a systematic procedure for describing and investigating 
finite-area topological defects, and, through this method, identify several of
these ``new" defects as divacancies and divacancy complexes.

\begin{figure}
\includegraphics[width=236pt]{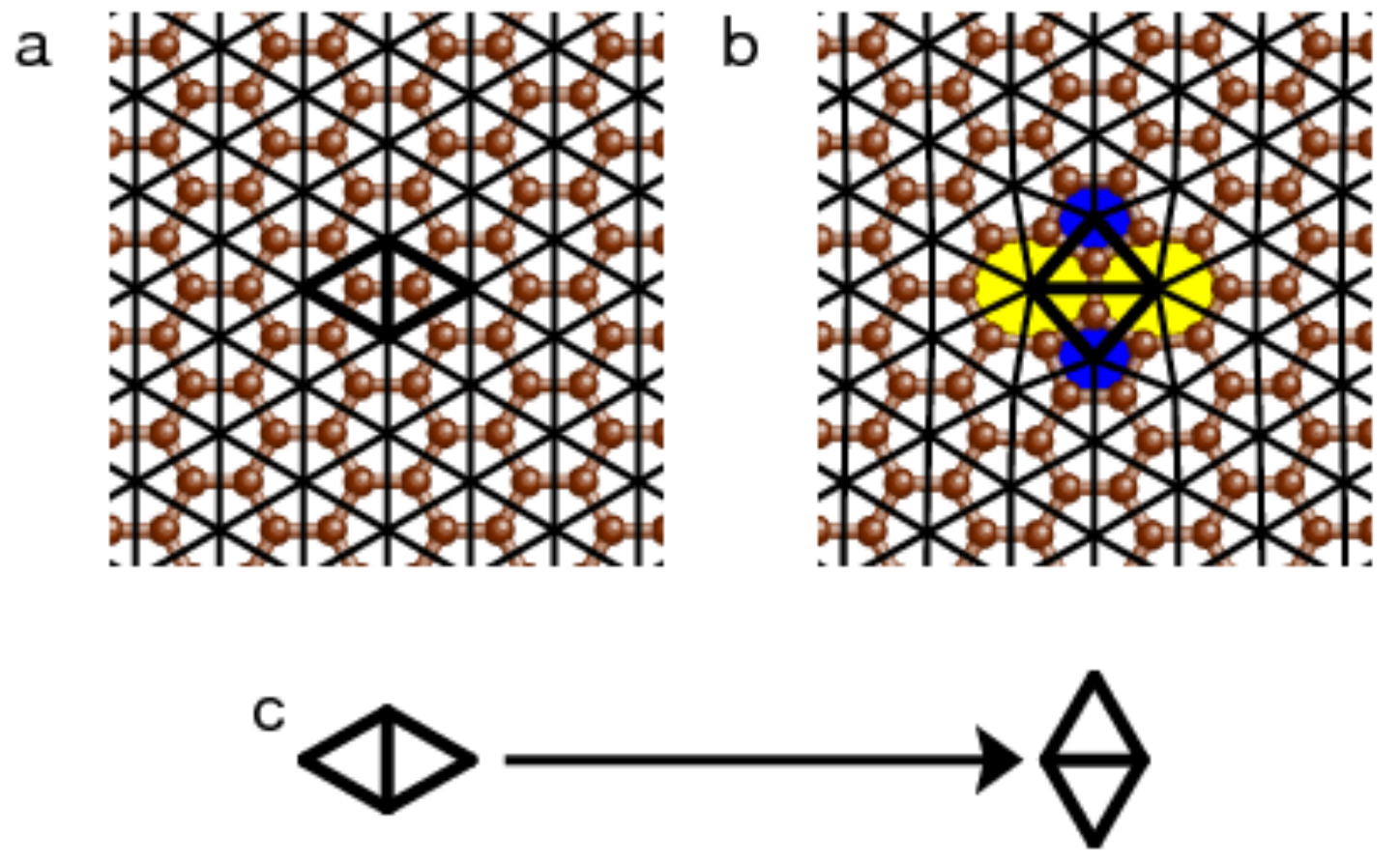}
\caption{(a) Ideal graphene honeycomb lattice with its dual tiling triangle tiling 
superimposed. (b) Graphene with Stone-Wales defect with its dual
tiling superimposed.  (c) Graphical 
representation of the Stone-Wales defect.}
\label{fig:dual}
\end{figure}

 The graphene structure can be represented as a honeycomb lattice
(\fig{dual}(a)).  Every planar lattice has a topologically 
equivalent {\em dual}\cite{Grunbaum86}, generated by converting
every $n$-vertex (vertex where $n$ edges meet) to an $n$-tile
(tile with $n$ sides), and vice versa.  Each edge in a tiling 
has exactly one corresponding edge in its dual.  The dual of the honeycomb
tiling is a regular triangular tiling (\fig{dual}(a)).
Similarly, every planar sp$^2$ bonded carbon structure
(three bonds per carbon) has a dual lattice that is composed
solely of triangles. The structure of topological defects in graphene
can thus be represented in terms of how they change a regular
triangular lattice into another triangular lattice.

 Consider for example, the Stone-Wales\cite{Stone86} defect (\fig{dual}(b)).  
It rotates a graphene C-C bond by 90$^\circ$.
Its dual-lattice equivalent is the rotation of
a patch of two joined triangles by 90$^\circ$ (\fig{dual}(c)).
The dual relationship between two lattices is mutual.
Therefore, it is possible to design a sp$^2$-type defect structure
in ``dual space" by replacing a chosen patch 
of the ideal triangle tiling with a different triangulated patch
of the same perimeter, and then taking its dual.
Metaphorically, one cuts a patch out of graphene and then 
``grafts" or ``transplants" a different sp$^2$-bonded patch with
the same number of dangling bonds.
Of particular interest is regions where the dual space replacement
patch is also a portion of an ideal triangular lattice. 
Because the replacement structure is graphitic in this
case, the formation energy should be relatively low.
In such cases, one effectively has a small grain of graphene inside bulk 
graphene with a closed grain boundary separating the regions.  
We called such defects ``grain boundary loops" in 
Ref.~\onlinecite{Cockayne11}.

\begin{figure}
\includegraphics[width=236pt]{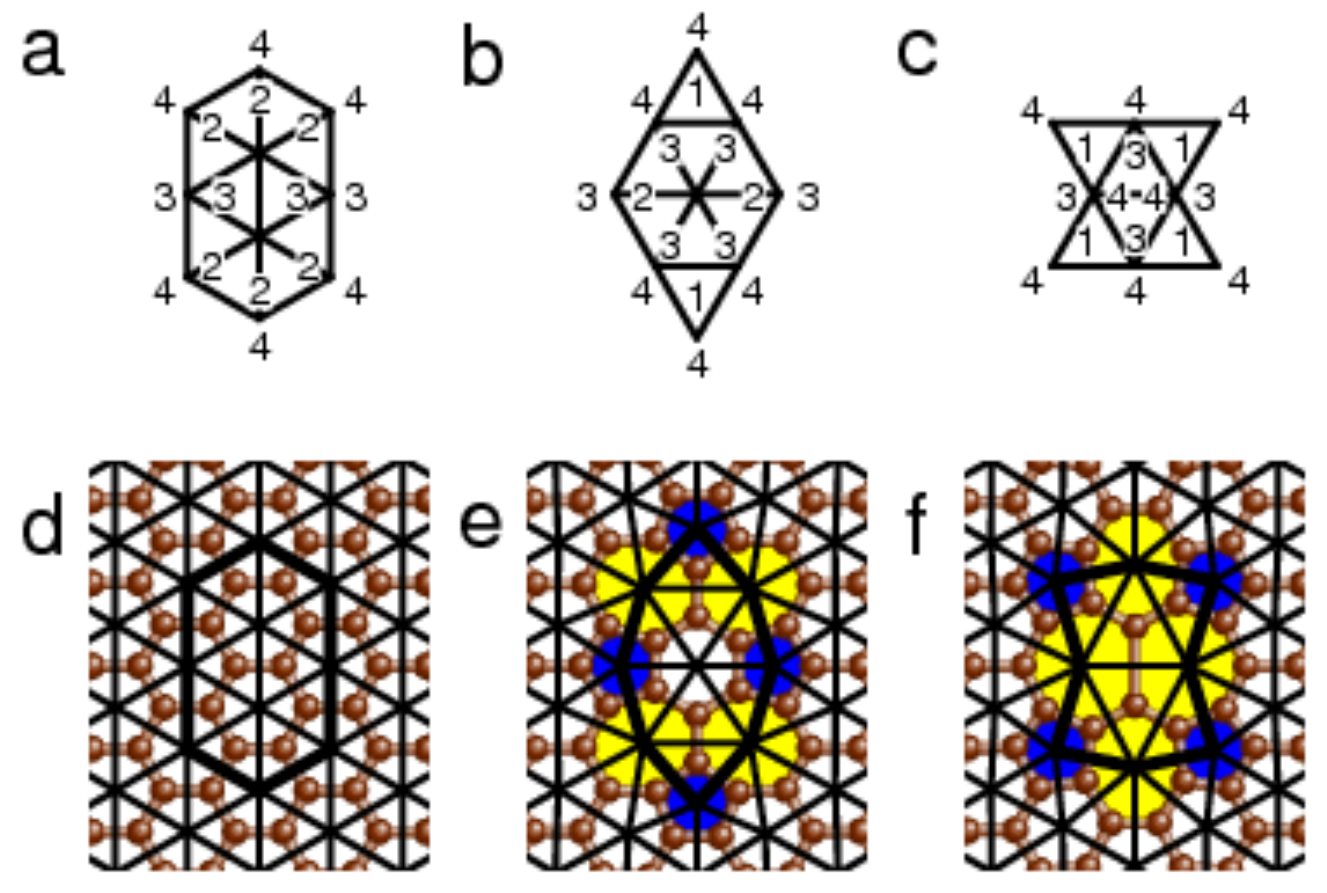}
\caption{(a) Patch of triangular tiling.  (b;c) Two replacement
patches, related by ``most compatible donor" procedure.  (d-f)
Corresponding graphene structures.  As described in the text,
and seen here, the number of triangles surrounding a given 
vertex in dual space (the sum of the interior and exterior 
counts shown) equals the perimeter of the corresponding carbon 
ring in the defect structure.}

\label{fig:transplant}
\end{figure}

 There are clearly an infinite number of possible grain boundary loops
in graphene.  In dual space, they are described by the shapes of the
(same-perimeter) triangular regions of the inner grain region
``before" and ``after" the procedure (\fig{transplant}), with the 
relative orientation an additional degree of freedom.
Note that the ``grafting" procedure provides a {\em description}
of various topological defects and not how these defects actually
form.  This description complements an alternative description of 
topological defects in terms of the sequence of elementary steps 
(divacancies, addimers, and Stone-Wales type bond rotations) that
could generate a given defect out of ideal graphene\cite{Appelhans10}.

Using the dual space formalism, it can easily be shown that the number
of dangling bonds $N_{db}$ in a graphene patch satisfies
$N_{db} = 3 N_{C} - 2 N_{fb}$, where $N_C$ is the number of carbon atoms
and $N_{fb}$ is the number of ``full" bonds in the interior.  As a 
corollary to this formula, $N_C$ must be even (odd) if $N_{db}$ is
even (odd).  Therefore, grain boundary loops, which preserve 
$N_{db}$, can only change the number of carbon atoms by a multiple of 2.

 A $n$-vertex in dual space corresponds to a $n$-ring in graphene,
that is, a ring of $n$ bonded carbon atoms.  The number of bonds emanating
from a vertex in dual space equals the number of triangles
surrounding that vertex.  The number of triangles surrounding a vertex
in the defect structure equals the number of exterior triangles sharing
that vertex in the ``before" state $N_{eb}$ plus the number of interior triangles
sharing that vertex in the ``after" state $N_{ia}$ (see \fig{transplant}).  
Since $N_{eb} = 6 - N_{ib}$, where $N_{ib}$ is the number of ``before"
internal triangles, $n = 6 - N_{ib} + N_{ia}$.

 Studies of energies of various fullerenes containing 5-rings and 7-rings
have led to the rules that the number of adjacent 5-5 pairs should
be minimized and adjacent 5-7 pairs maximized\cite{Ayuela96}.  
A similar empirical rule has been noted for topological defects in 
graphene\cite{Cockayne11,Kotakoski11}: it is energetically favorable for grain
boundary loops to consist of alternating 5-rings and 7-rings.
In terms of the above formula, alternating 5-rings and 7-rings are obtained
if $N_{ia} - N_{ib}$ is alternately +1 and -1 for the vertices on the 
perimeter of the replacement triangular patch in dual space.  
Continuing the transplant metaphor, we define 
``most compatible donor(s)" for each patch of the ideal triangle 
lattice (if any exist), as those patches whose perimeter vertices alternately 
span $N_{ia} \pm 1$ and $N_{ia} \mp 1$ triangles with respect to the
perimeter vertices of the original patch.  Because either choice of 
sign is possible, a structure can have up to two inequivalent 
most compatible donors.
\fig{transplant} shows the two most compatible donors for one
representative patch.  The resultant defects are distinct; 
one reduces the number of carbon atoms by 2 and the other reduces it by 4.

 We hypothesize that the most significant finite-area topological defects
in graphene are grain boundary loops  described by the 
``most compatible donor" procedure.  We have enumerated all such defects 
with $C_{2v}$ or higher symmetry that are contained within a diameter of
1.0 nm or smaller, and that change the number of carbon atoms by two or fewer.
We then investigated their stabilities and predicted their scanning tunneling
microscopy (STM) signatures using density functional theory.


First principles density functional theory (DFT) calculations, as encoded in
the Vienna ab-initio Simulation Package (VASP) software\cite{Kresse96}, 
were used perform relaxations and to calculate the local density of states for STM simulations.
Vanderbilt-type ultrasoft pseudopotentials\cite{Vanderbilt90}
were used with a plane wave basis set with a cutoff energy of 211.1 eV. 
Calculations were mostly performed on supercells
with approximately 216 and 486 atoms.  The results were extrapolated
to estimate formation energies for isolated defects and their
uncertainties.
Efficient Brillouin zone integration was performed using a
mesh containing six k-points in each supercell Brillouin zone
((1/3,0,0) and symmetry equivalents).  These parameters were sufficient
to obtain formation energies in good agreement with previous
results\cite{Banhart11}.  
Further methodological details are given in \cite{Cockayne11}.

The in-plane lattice parameter was set to the value for graphite,
corresponding to epitaxial multilayer growth conditions.
Structures with more atoms than ideal graphene were allowed to
relax out-of-plane, while structures with the same number of atoms 
or fewer atoms were fixed to remain planar.  There are reports that
such defects may prefer to relax out of plane\cite{Ma09}. 
However, weak buckling instabilities might be suppressed
under experimental conditions
by adhesion of the surface layer in few-layer graphene to
to the layer below. This is an issue for further
exploration.

STM images were simulated for comparison with experimental 
fixed-voltage topographs.
The tunneling current is approximated as proportional to the
local electronic density of states integrated between $E_F$ and
$E_F + e  {\Delta}V$, with $E_F$ the Fermi level and ${\Delta}V$ the 
bias voltage\cite{Tersoff85}.
For these calculations, $4 \times 4 \times 1$ grids of k-points in 
the Brillouin zone supercell were used, centered at the origin.
Using the experimental value of $e {\Delta}V = + 0.3~{\rm eV}$, the best 
agreement with experimental STM images in \cite{Rutter07} comes 
from setting $E_F = E_D + 0.05~{\rm eV}$ ($E_D$ the Dirac level)  rather
than the experimental value $E_F = E_D + 0.3~{\rm eV}$.  A significant
part of the discrepancy comes from simulating a 
monolayer rather than the experimental case of Bernal-stacked 
few-layer graphene. Coupling between graphene layers in 
Bernal-stacked bilayer graphene raises the energy of the low-lying 
antibonding states of the carbon atoms on the stacked sublattice by about
0.22 eV\cite{Castro07}.  Aside from the
offset discrepancy, very good agreement between the monolayer simulation and
the few-layer experiment is found with greatly reduced computational
cost compared with multilayer DFT calculations.


\begin{figure}
\includegraphics[width=236pt]{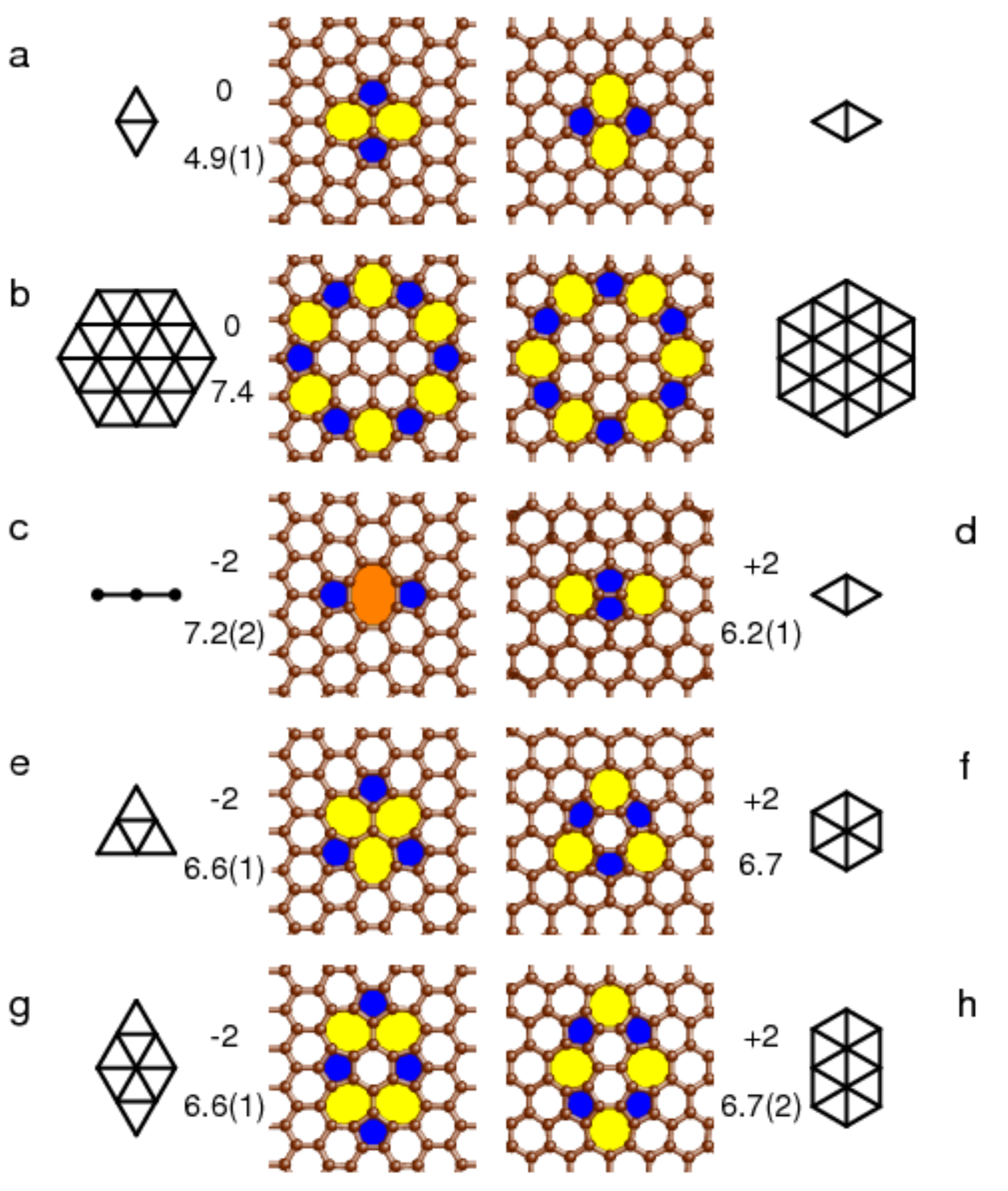}
\caption{Small high-symmetry topological defects in graphene:
(a) Stone-Wales\cite{Stone86}. (b) Flower\cite{Park10}. 
(c) Divacancy. (d) Addimer.  
(e) $V_2$ (555-777)\cite{Lee05,Banhart11}. 
(f) H$_{5,6,7}$\cite{Qi10}. (g) $V_2$(5555-6-7777)\cite{Banhart11}.
(h) 2-hexagon addimer reconstruction\cite{Orlikowski99}.  
Defects related by inversion are side-by-side.  Opposite and 
same-side triangle patch are the before-and-after representation 
in dual space.  Next to the relaxed structures are the change in number of
atoms (above) and the formation energy (below), in eV, with 
uncertainty in units of 0.1 eV in parentheses (if greater than 0.05 eV).}
\label{fig:struc}
\end{figure}

\begin{figure}
\includegraphics[width=236pt]{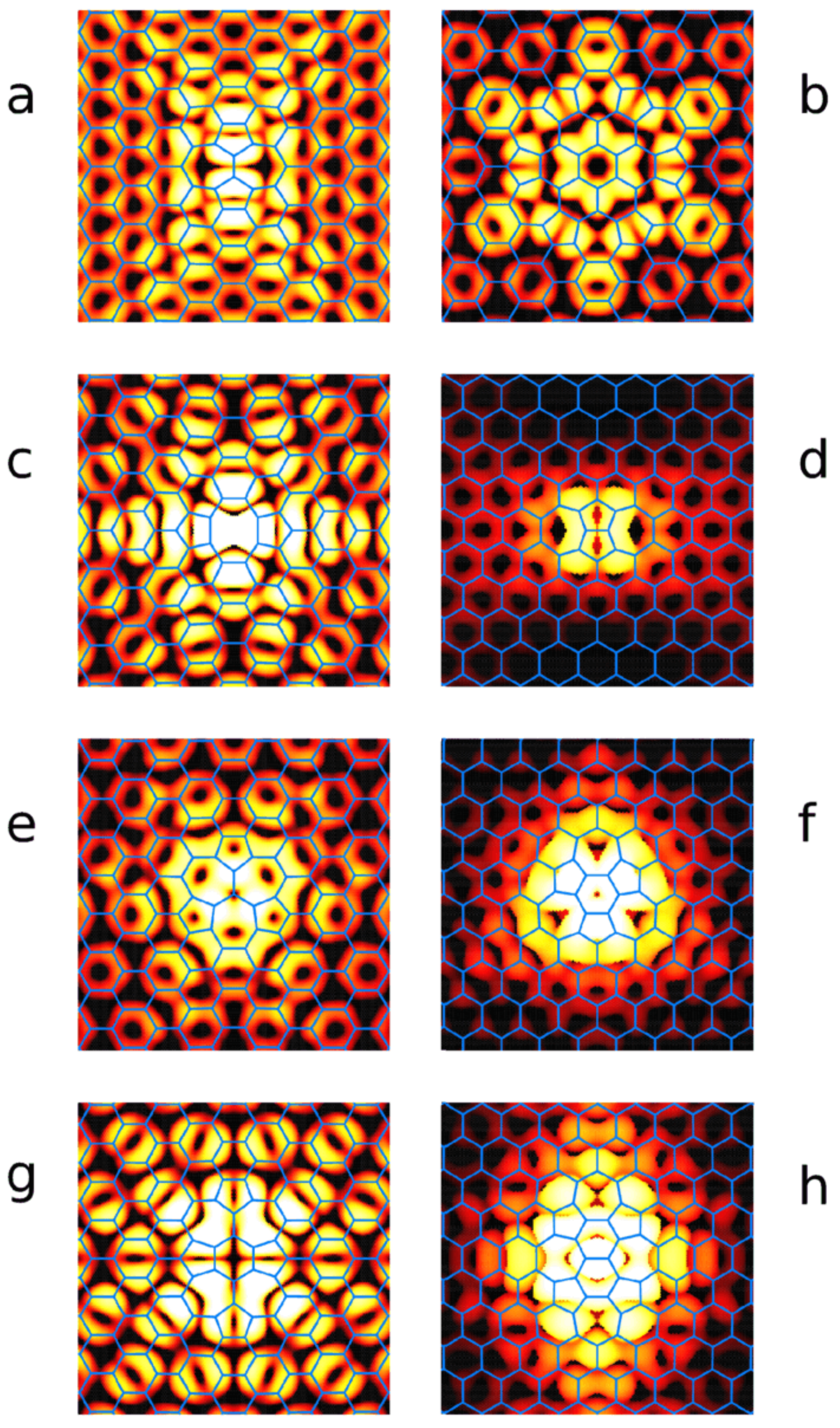}
\caption{(a-h) Simulated STM topographs for the defects shown in 
\fig{struc}~(a-h), respectively.  Height range from brightest to darkest
is 0.32 nm in nonplanar defects (d),(f), and (h), and
0.12 nm otherwise.  All images in \fig{stm} to \fig{zzz} show 1.0 $\times$
1.0 nm patches centered on a defect.}
\label{fig:stm}
\end{figure}

 The various small high-symmetry graphene defects are shown in \fig{struc}.
The figures show the graphical representation of each defect,
its relaxed atomic structure, the change in the number of atoms 
with respect to ideal graphene, and the formation energy.
The structures that increase the number of atoms all have out-of-plane
distortions of order 0.2 nm to 0.3 nm.  Their relaxed structures have
the same ``rounded hill" appearance shown for such defects in
the literature\cite{Lusk08}, and are not shown here.
The corresponding STM simulations of the defects are shown in
\fig{stm}. The height scale is larger for those structures that
distort out-of-plane.

\label{sec:discussion}

The defects shown side-by-side in \fig{struc} have graphical
representations that are the inverse of each other,
except for the Stone-Wales and flower defects, which are
equivalent to their own inverses (as an aside, the
addimer defect, frequently termed the ``inverse
Stone-Wales defect\cite{Lusk08}", is not technically the inverse
of the Stone-Wales defect, but the inverse of the divacancy).
Using the dual formalism, it is easily seen that if a 
topological defect adds $n$ atoms, its inverse removes 
$n$ atoms.  

While the most compatible donor procedure (\fig{transplant})
was designed to yield structure with alternating 5 and
7 rings, in the case of the divacancy, a patch of two
triangles in dual space collapses into 
what is effectively a zero-area polygon with 4 sides
(\fig{struc}(c)).  The resultant structure has a new
polygon, namely an 8-ring, at the center.

The most compatible donor procedure 
yields dual structure patches whose triangles are
rotated $30^{\circ}$ with respect to those of the original
dual.  When the patches have sufficiently high symmetry, strain
mismatch is minimized when the transplant patch is oriented
{\em exactly} $30^{\circ}$ with respect to the surrounding
structure.
The procedure in this work thus explains the experimental
observation of of $30^{\circ}$ rotated regions inside grain boundary
loops\cite{Kotakoski11}.

Two divacancy reconstructions, $V_2 (555-777)$
and $V_2 (5555-6-7777)$ are found that are lower energy
than the simple divacancy.  Such energy-lowering 
reconstructions have previously been 
reported\cite{Lee05,Banhart11}.  In contrast to \cite{Banhart11},
the $V_2 (5555-6-7777)$ defect energy is lower than
the $V_2 (555-777)$ one for both periodic cells investigated,
although the difference between the extrapolated energies is less than
their uncertainties. Since the energy of a grain boundary loop tends
to increase with its perimeter\cite{Cockayne11}, it is unlikely
that there is any undiscovered divacancy reconstruction that
has lower energy (unless it has lower symmetry than explored here).

For an addition of two atoms, the addimer defect has lowest energy.
Reconstructing the addimer gives configurations that are higher
in energy.  Note, however, that both addimer reconstructions
in \fig{struc} have appeared in numerical simulations of addimers on
carbon nanotubes under strain\cite{Orlikowski99}.

Topological defects can be combined to make more complex defect
structures.  In the dual space description nonoverlapping
regions are simultaneously retriangulated.
It is an open question whether and under which circumstances
various topological defects attract or repel, but it is interesting
that clustering has been observed experimentally
for flower defects\cite{Guisinger08} and for 
divacancies\cite{Kotakoski11}.

\begin{figure}
\includegraphics[width=236pt]{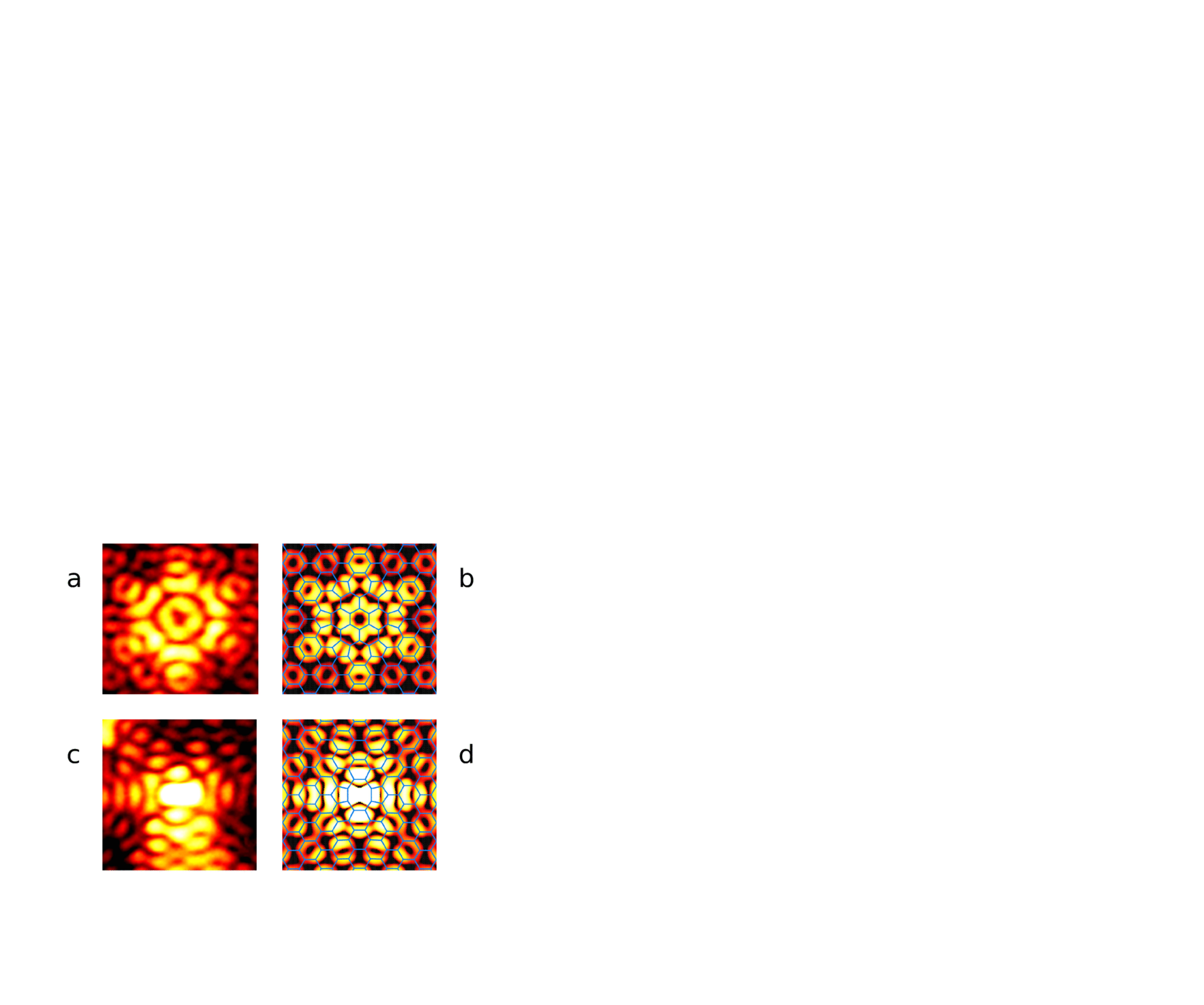}
\caption{(a) Experimental STM images of sixfold defect 
in graphene (all experimental images shown in 
\fig{xxx}-\fig{zzz} are adapted from
Ref.~\cite{Rutter07}). (b) Simulated STM image of
flower defect is a good match for (a).  
(c) Experimental STM image of small twofold defect
in graphene matches (d) simulated STM image of 
divacancy}.
\label{fig:xxx}
\end{figure}

 \fig{xxx} shows two defects observed in experimental
STM topographs\cite{Rutter07} that are well-matched by
simulations of topological defects shown in
\fig{stm}.  The sixfold defect shown in \fig{xxx}(a) 
agrees with the simulation of the flower 
defect (\fig{xxx}(b)), as previously shown 
in~\cite{Cockayne11}.  The twofold symmetric defect in
\fig{xxx}(c) agrees very well with the simulation of the
divacancy (\fig{xxx}(d)).  The characteristic dumbbell
shape of the central region was previously predicted by
Amara {\it et al.}\cite{Amara07}, and distinguishes
the STM image of this defect from that of the twofold symmetric
Stone-Wales defect.

\begin{figure}
\includegraphics[width=236pt]{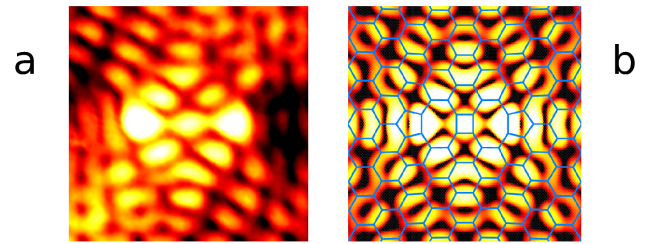}
\caption{(a) Experimental STM image of large twofold defect in
graphene. (b) The simulated STM image of a double divacancy
agrees well with (a).}
\label{fig:yyy}
\end{figure}

 A second twofold defect observed in \cite{Rutter07} is shown
in \fig{yyy}(a).  The double divacancy shown in \fig{yyy}(b)
is an excellent match to the experimental image.
The double divacancy has also been observed in
single-layer graphene created by mechanical cleavage and then
irradiated\cite{Kotakoski11}.  It contains a rectangle of 4 carbon atoms 
in its center.  Note that the simulated STM image is not
simply a superposition of the STM images of two individual
divacancies.

\begin{figure}
\includegraphics[width=236pt]{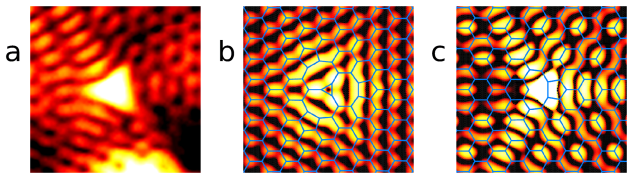}
\caption{(a) Experimental STM image of threefold defect in
graphene. (b) Simulated STM image of a triple divacancy matches
central concave region well but matches poorly outside this region.
(c) Simulated local density of states of  pseudo-threefold vacancy
between E$_F$ + 0.1 eV and E$_F$ + 0.2 eV matches peaks and nodes
of experimental image very well}.
\label{fig:zzz}
\end{figure}

 Finally, \fig{zzz}(a) presents the experimental STM image of a
defect with threefold symmetry.  This defect was initially modelled
as a triple divacancy (\fig{zzz}(b)) in analogy with the double
divacancy. The triple divacancy has a triangle of carbon atoms in the middle.
While a planar triangular carbon configuration is perhaps unexpected on
energetic grounds, a C-C-C triangle naturally occurs
for the low-energy nonplanar bridging configuration of a single
carbon adatom on graphene\cite{Maiti97,Li05}, and 
a C-C-C triangle also occurs in the cyclopropane molecule.  
Furthermore, the calculated formation energy of a triple divacancy in the configuration
shown is 15.9 $\pm$ 0.5 eV, less than three times the formation energy of
three isolated divacancies, 7.2 $\pm$ 0.2 eV each (\fig{struc}),
giving a plausibility argument for the arrangement in
\fig{zzz}(b).  (The calculated energy of the double divacancy, 11.5 $\pm$ 0.9 eV is also less
than the total for the isolated divacancies, further showing the tendency of divacancies to
cluster in graphene.) The simulated STM image in \fig{zzz}(b) reproduces 
the central concave triangle seen experimentally.  
The peaks and nodes outside this region, however, are poorly reproduced.

 An alternate explanation for the experimental threefold defect is a single
vacancy (\fig{zzz}(c)).  Although a vacancy in graphene relaxes in a way that
breaks threefold symmetry\cite{ElBarbary03,Amara07}, the calculated local 
density of states between E$_D$ + 0.1 eV and E$_D$ + 0.2 eV is found to have pseudo-threefold 
symmetry (\fig{zzz}(c)).  (The DFT calculations for this simulation included
spin polarization to incorporate magnetism.)
The electronic structure in this energy range reproduces the concave central
region and furthermore, the peak and node structure outside this region matches
that of the experimental STM image.  The simulated image in \fig{zzz}(c)) looks almost
exactly like that simulated for a generic impurity on the surface of graphite
by Mizes and Foster\cite{Mizes89}.  Their qualitative model was based on a 
tight-binding approximation and is equally valid for an adatom directly above
a carbon, a vacancy, or a substitution.  It is concluded that the threefold
experimental defect is one of these three types that is located on a single 
carbon site.  A problem with interpreting the defect as a vacancy is
that, when simulated to match the experimental bias of 0.3 eV, either from
E$_D$ + 0.05 eV and E$_D$ + 0.35 eV as for the other STM simulations, 
or for any other 0.3 eV range near $E_D$, then the agreement with experiment is lost.
A possible explanation for the discrepancy is strong finite-size unit cell artifacts, as 
the perturbation due to a vacancy falls off as $1/r$\cite{Bena08}, more 
slowly than for topological defects that preserve sp$^2$ binding.
STM images for a single vacancy in graphene need to be simulated for larger 
unit cells to test this hypothesis and compared with similar simulations for adatoms
and substitutions.





In conclusion, we introduce a powerful dual space method for describing
finite-area topological defects in graphene.  This method allows us
to systematically explore candidate low-energy defects and defect clusters.
Previously unidentified defects in graphene were identified as
divacancies or divacancy clusters.



Acknowledgements: I thank O. Yazyev for helpful discussions and J. Stroscio for assistance
with experimental figures.

\end{document}